\documentclass[12pt,preprint]{aastex}

\shorttitle{Absence of Electron Surfing Acceleration}
\shortauthors{Ohira and Takahara}

\begin{document}

\title{Absence of Electron Surfing Acceleration in a Two-Dimensional Simulation}

\author{Yutaka Ohira and Fumio Takahara}
\affil{Department of Earth and Space Science, Graduate School
of Science, Osaka University, 1-1 Machikaneyama-cho, Toyonaka, Osaka
560-0043, Japan \\
{\rm yutaka@vega.ess.sci.osaka-u.ac.jp, takahara@vega.ess.sci.osaka-u.ac.jp}}

\begin{abstract}
Electron acceleration in high Mach number perpendicular shocks is investigated through two-dimensional electrostatic particle-in-cell (PIC) simulation. We simulate the shock foot region by modeling particles that consist of three components such as incident protons and electrons and reflected protons in the initial state which satisfies the Buneman instability condition. In contrast to previous one-dimensional simulations in which strong surfing acceleration is realized, we find that surfing acceleration does not occur in two-dimensional simulation. This is because excited electrostatic potentials have a two-dimensional structure that makes electron trapping impossible. Thus, the surfing acceleration does not work either in itself or as an injection mechanism for the diffusive shock acceleration. We briefly discuss implications of the present results on the electron heating and acceleration by shocks in supernova remnants. 
\end{abstract}

\keywords{supernova remnants -- shock waves -- plasmas -- 
cosmic rays -- acceleration of particle}

\section{Introduction}
The discovery of synchrotron X-rays from young supernova remnants (SNRs) provides the evidence that electrons are accelerated to highly relativistic energy in SNR shocks \citep{koy95}. Since SNR shocks are collisionless, various processes of particle acceleration may take place. Although the most popular acceleration mechanism is diffusive shock acceleration \citep{bel78,bla78}, electrons are not expected to be accelerated easily because they behave adiabatically in the field made by ions in the first order approximation. Thus, in order for electrons to enter the acceleration process they should be pre-accelerated to relativistic energy by some injection mechanisms. Surfing acceleration at perpendicular shocks has been studied and regarded as one such prospective mechanism, as well as an efficient direct acceleration mechanism \citep{hos02,mcc01}.

The basic mechanism of electron surfing acceleration for high Alfv${\rm \acute{e}}$n Mach number perpendicular shocks is as follows. The shock reflects some of the incident ions to the upstream, where the foot region forms on a spatial scale of ion gyroradius \citep{ler83}. The plasma in the foot region consists of incident ions and electrons and reflected ions. \citet{pap88} proposed that in this situation, incident electrons and reflected ions excite electrostatic waves by the Buneman instability \citep{bun58} because the relative velocity between them is fairly large compared with the electron thermal velocity and because the waves heat electrons non-adiabatically. \citet{car88} performed a one-dimensional hybrid simulation and demonstrated that strong electron heating actually does occur. 

To examine the electron momentum distribution, \citet{shi00} performed one-dimensional full-PIC simulations in which both electromagnetic and electrostatic waves were taken into account but with a small proton/electron mass ratio. Their results showed that significant electron surfing acceleration occurs, which can be explained in the following way: in the shock rest frame, electrostatic waves excited by the Buneman instability propagate away from the shock front with a large amplitude in order to trap some fraction of electrons that are accelerated in the direction of $\mbox{\boldmath $V$}_{\rm sh}\times \mbox{\boldmath $B$}$ by the motional electric field ($\mbox{\boldmath $V$}_{\rm sh}$ is the shock velocity and \mbox{\boldmath B} is the magnetic field). Since the potential structure is uniform in the direction of this acceleration in one-dimensional simulations, electrons are kept to be trapped and accelerated indefinitely. The degree of acceleration is determined by the coherence length of the potential structure to the acceleration direction and the central question is whether it is long enough for real shocks in SNRs.

In this Letter, we address this question by using a two-dimensional electrostatic PIC simulation. In \S 2 we describe the setting and results, followed by a discussion in \S 3.  

\section{Simulation}
To perform two-dimensional simulations, we confine our attention to the foot region through a proper modeling instead of solving the whole shock structure. Our simulation box is taken to be at rest in the upstream frame of reference, i.e., that of incident protons and electrons. We do not solve electromagnetic waves; we concentrate on electrostatic waves and adopt a real proton/electron mass ratio of 1836.

\subsection{setting}
We define the $x$-direction as the shock normal pointing to the shock front, and thus the reflected protons move in the $-x$-direction. The magnetic field is taken to be spatially homogeneous pointing in the $z$-direction and we solve the particle motion and electric field in the $x-y$ plane. As the initial condition, we prepare upstream electrons, upstream protons and reflected protons. Each population is uniformly distributed in the $x-y$ plane and their momentum distribution is given by a Maxwellian at the same temperatures $T=T_e=T_p=T_r=7{\rm eV}$. In addition, reflected protons have an extra drift velocity in the $x$-direction of $V_d=-0.04c$ ($V_d=2V_{\rm sh}$). Number densities of each population are taken as $n_e=1.25n_p=5n_r=1\rm{cm}^{-3}$, where subscripts e, p and r represent upstream electrons, upstream protons and reflected protons, respectively.  These parameters are typical of young SNRs. 
 
We employ the periodic boundary condition in both the $x$- and $y$-directions. The electric field is solved by a Poisson equation. We have examined various values from 0 to 72 $\mu $G for magnetic field strength ($\omega_{ce}/\omega_{pe}=0\sim0.08$), where $\omega_{ce}=eB/m_{e} c$ and $\omega_{pe} = (4\pi n_{e} e^2/m_{e})^{1/2}$ are electron cyclotron frequency and electron plasma frequency, respectively. Among them, in this Letter, we present only one specific case for which the maximum energy of accelerated electrons is largest in the one-dimensional (1D) simulation; i.e., the magnetic field strength is 27$\mu$G ($\omega_{ce}/\omega_{pe}$=0.03). 
 
The size of the simulation box to the $x$- and $y$-directions is taken to be $L_x=L_y=16\lambda_{\rm Bun}$, with a total of 256 $\times$ 256 cells, where $\lambda_{\rm Bun}=2\pi \omega_{pe}/V_d$ is the wavelength of the most unstable mode of the Buneman instability.  Thus, the length of each cell $\Delta x$$=\Delta y$ is 3 times the Debye length.  We also perform 1D simulations for the purpose of comparison with two-dimensional (2D) simulations. The number of macroparticles is taken so that initially each cell includes 10,240 and 80 electrons in the 1D and 2D cases, respectively. Time step $\Delta t$ is taken as 0.01$\omega_{\rm pe}^{-1}$ and the simulation is followed until 1000$\omega_{\rm pe}^{-1}$.

\subsection{behavior of the electric field}

First, we discuss the time development of the electric field. The energy density of the electric field is shown in Figure 1. The three curves show the spatially averaged energy density of the electric field. The solid curve shows the 1D simulation, while the dashed and dotted curves show the $x$- and $y$-components in the 2D simulation, respectively. These three curves show a comparable growth rate although the growth in the 2D case appears to start earlier, probably due to a smaller number of macroparticles per cell used and larger initial fluctuations. It reaches maximum at about a hundred times $\omega_{pe}^{-1}$, with an $e$-folding time of about 10$\omega_{pe}^{-1}$, as the linear theory predicts. It should be noted that the amplitude of the $y$-component is comparable to that of the $x$-component, which is discussed below. The electric field strength for 2D simulation is a few times smaller than that of the 1D case, and its time history is far smoother. This is because the phase of electron plasma oscillation is less coherent in the 2D simulations than that in the 1D case. Two-dimensionality of electric field structure is clearly seen in a landscape representation of the electrostatic potential at $t=150 \omega_{pe}^{-1}$, as shown in Figure 2. The typical wavelength in the  $x$-direction is $\lambda_{\rm Bun}$, while that in the $y$-direction shows a larger dispersion and the typical coherence scale of the potential fluctuations in the $y$-direction $L_{\rm coh}$ turns out to be on the order of $\lambda_{\rm Bun}$, too. 

The fact that obliquely propagating modes have a comparable growth rate with parallel propagating modes was pointed out as early as 1960 by \citet{blu60}, who derived the dispersion relation against obliquely propagating electrostatic two-stream instability for cold beams without a magnetic field. \citet{lam74} discussed thermal effects and made numerical simulations to study nonlinear evolution for initially cold unmagnetized plasma and a somewhat smaller ion-electron mass ratio. Because the magnetic field is sufficiently weak in our simulations, we may safely ignore effects of the magnetic field on the growth timescale of the most unstable mode. The dispersion relation for the present case, ignoring thermal effects, is written as 
\begin{equation} 
0=1-\left(\frac{\omega_{pe}}{\omega}\right)^2
   -\left(\frac{\omega_{pr}}{\omega -k_x V_{d}}\right)^2,
\end{equation}
where $\omega_{pr}=(4\pi n_r e^2/m_p)^{1/2}$ and we safely ignored the effect of upstream protons and the relativistic effect. The wavenumber in the $x$-direction of the most unstable mode is given by 
\begin{equation}
k_x = \frac{2\pi}{\lambda_{\rm Bun}}.
\end{equation} 
Note that equation (1) is the same as the dispersion relation derived by Buneman for a 1D case and does not contain $k_y$ \citep{bun58}. This means that if $k_x$ satisfies the instability condition, the system is unstable against all values of $k_y$ for cold beams, as previous studies have shown. \citet{lam74} showed that as electrons are rapidly heated, waves with $k_y$ comparable with $\lambda_{\rm Bun}^{-1}$ survive. Our 2D simulation results are fully consistent with their results; growth of the $y$-component of the electric field is a natural feature of the two-stream electrostatic instability. We note that this natural feature is not taken into account in 1D simulations that discuss the surfing acceleration. 

\subsection{momentum distribution of electrons}

Momentum distribution of electrons in the 2D simulations is also different from that in the 1D case. Figure 3 shows the spatially averaged momentum distribution of electrons at $t= 690\omega_{\rm pe}^{-1}$. We find that in the 1D case some electrons are accelerated to the $-V_y$ direction (and subsequently rotate counterclockwise), as seen in several jetlike features in the velocity space; these are identified with the surfing acceleration. Therefore, we confirm that surfing acceleration occurs in the 1D simulation as has been discussed in the literature. In contrast, in the 2D case, we do not see any jetlike structures as seen in the 1D case. The distribution is concentrated in a central round region and thus no surfing acceleration occurs.

Figure 4 shows the energy spectrum of electrons at the end of the simulations. A high-energy component representing the surfing acceleration is seen in the 1D case, while no high energy components are seen in the 2D case. The energy distribution is smooth but is not described as Maxwellian with a single temperature. In Figure 4, the distribution for the 2D simulation without a magnetic field is also shown for comparison. The result with a magnetic field is somewhat hotter than that with no magnetic field. A question may be raised as to whether the heating increases with magnetic field strength and whether we get sufficient acceleration for a sufficiently strong magnetic field. To see this, we estimate the effect of the magnetic field on electron energy change over a coherence scale of potential $L_{\rm coh}$, as
\begin{equation}
\Delta E =e\frac{B}{c}V_d L_{\rm coh}
= 2\pi \alpha m_e V_d^2 
  \left(\frac{\omega_{ce}}{\omega_{pe}}\right),
\end{equation}
where we set 
\begin{equation}
L_{\rm coh}=\alpha \lambda_{\rm Bun},
\end{equation}
with $\alpha$ being of the order of 0(1). This is an order of magnitude too small to explain the difference in the simulations because $\omega_{ce}/\omega_{pe}$ is 0.03. When we inspect the time history of electron distribution, we observe that the difference appears after the electric-field strength begins to decay. So we think that the difference is caused by the thermalization process after the saturation of electric field fluctuations. Actually, our simulation results for various magnetic field strengths do not show a significant difference in the energy distribution. Therefore, we conclude that surfing acceleration does not work in the 2D simulations or in real cases. 

\subsection{interpretations}
In essence, in realistic 2D cases, the coherence length of waves excited by the Buneman instability is the order of $\lambda_{\rm Bun}$ in both the $x$- and $y$-directions, with a negligible contribution of very long wavelength modes required for surfing acceleration. Our simulations do not show stochastic surfing acceleration mentioned by \citet{mcc01}, either. This is because the stabilization of finite amplitude waves by BGK mode does not work in the 2D case; there are no large-amplitude waves, since electrons leave the potential on the timescale of the electron cyclotron period. Even when they are trapped in the $x$-direction, they leave it to the $y$-direction, which hampers nonlinear growth of the waves. 

Thus, the net result is not the acceleration but the heating, as was discussed in the original investigations of Papadopoulos (1988). The electron temperature in the  final state in our 2D simulation, inferred from the obtained velocity dispersion, is about $T_f\sim m_e V_ d^2$, the same as the result of \citet{lam74} without a magnetic field. Proton temperature does not change much during the simulation because protons react on a much longer timescale. This temperature corresponds to the threshold temperature for the Buneman instability. As was shown in \S 2.3, electric-field amplitude decreases after it attains the maximum. This does not mean the electric field energy is simply transformed into the electron thermal energy. The former is much smaller than the latter. The electric field plays a catalytic role in transforming the kinetic energy of reflected protons. As the electron temperature increases, the growth of the electric field is suppressed, and when it attains the threshold temperature, the wave growth ceases. The detailed examinations of the heating process are interesting but beyond the scope of this Letter (see \citet{lam74}). 

\section{Discussion}
The present study has made some assumptions that we believe are not so crucial as to invalidate the physics conclusions. First, we have limited ourselves to only electrostatic modes and do not solve electromagnetic waves. This is sufficient in the linear regime but may influence nonlinear evolution of the instabilities. Although the two-stream situation is known to be also unstable against electromagnetic modes, their growth rate is orders of magnitude smaller than that of electrostatic modes. In the late stage when the electrostatic modes saturate and begin to decay, electromagnetic modes may play an important role in electron heating and acceleration. The behavior will surely depend on the electron temperature attained in the initial stage dominated by the electrostatic modes. 

In this Letter, we have presented the result of only one specific parameter set. The results for other parameter sets will be discussed elsewhere. Here, we briefly discuss effects of the initial electron temperature. We have discussed the case in which the drift velocity is about 7 times the thermal velocity of the electrons ($T=7$eV). We have performed simulations for other values of the initial electron temperature, keeping the drift velocity unchanged. When the temperature is 4 times lower, i.e., $T=1.75$eV, the structure of potential in the $y$-direction becomes more fluctuated because modes with shorter wavelengths in the $y$-direction can grow, consistent with our argument in \S 2.2. Conversely, when the initial temperature is higher, growth of short-wavelength modes is suppressed, and long-wavelength modes preferentially survive; thus, one may expect that more heating or even surfing acceleration may be realized. In reality, this is not the case. We have done simulations for which the temperature is 4 and 16 times higher, i.e., $T=28$ and 112 eV. The resultant coherence length of potential in the $y$-direction becomes longer, as expected, but it is only a few times longer and does not much change the situation. It should also be noted that when the temperature is higher, potential amplitude becomes small and less efficient in trapping electrons because the thermal effect becomes significant. When the initial temperature is as high as the one corresponding to the drift velocity, the Buneman instability does not develop. Thus, the attained temperature by the instability remains of the same order. The conclusion that surfing acceleration does not work is also valid. 

Our simulations do not treat three-dimensional effects. As long as the magnetic field is weak, the $y$- and $z$-directions are not so much discriminated, and we expect that our conclusions remain safe. When the magnetic field is extremely large, the situation may change, but at the same time the Alfv${\rm \acute{e}}$n Mach number becomes small and the shock will be weaker. 

Our results have several implications for the electron heating and acceleration in SNR shocks. The attained electron temperature in simulations is comparable to the observed temperature of the thermal X-ray components from young SNRs. As electrons pass the front to downstream, they suffer further adiabatic heating, which may explain the small difference between the attained and observed temperatures, in addition to the Coulomb heating. This in turn may suggest that the degree of further heating in the upstream such as that due to the excitation and decay of ion acoustic modes is not significant. This point deserves further study. As for the acceleration of nonthermal electrons, the question remains open. Since the attained temperature is on the order of kilo-electron volts, it is still 4 orders of magnitude short of the energy necessary to enable injection into the diffusive shock acceleration.

\section{Summary}
We performed electrostatic 2D PIC simulations to investigate electron heating and acceleration by collisionless shocks with a high Mach number. We consider only the foot region by properly modeling the effects of reflected protons. We have shown that excited electrostatic waves are oblique to the shock normal, and the electrostatic potential loses coherence in the direction perpendicular to the shock normal and the magnetic field. As a result, surfing acceleration seen in the previous 1D simulations does not occur in 2D simulations or in real cases. 

\acknowledgments
We are grateful to T. Tsuribe, T. Kato and Y. Fujita for discussions and suggestions especially for providing useful guidance and information in doing numerical simulations.

\clearpage

\begin{figure}
\plotone{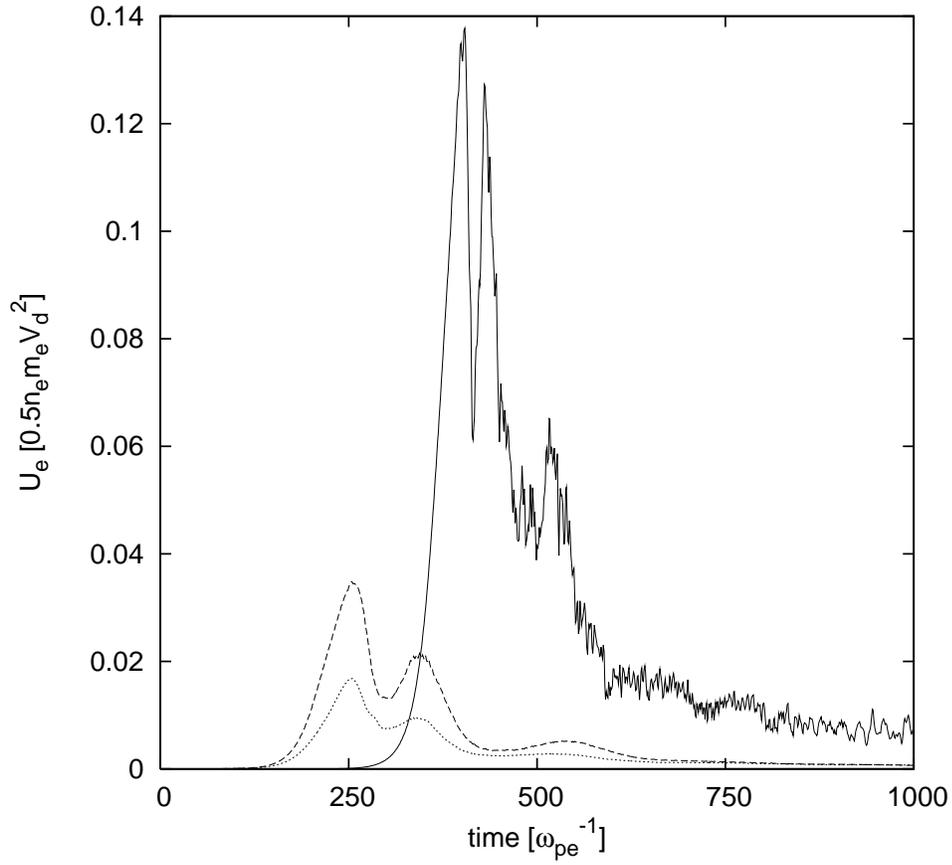} \caption{Time development of the spatially averaged energy density of the electric field. Solid, dashed, and dotted curves represent the 1D simulation and the $x$- and $y$-components in the 2D case, respectively. \label{fig1}}
\end{figure}


\begin{figure}
\plotone{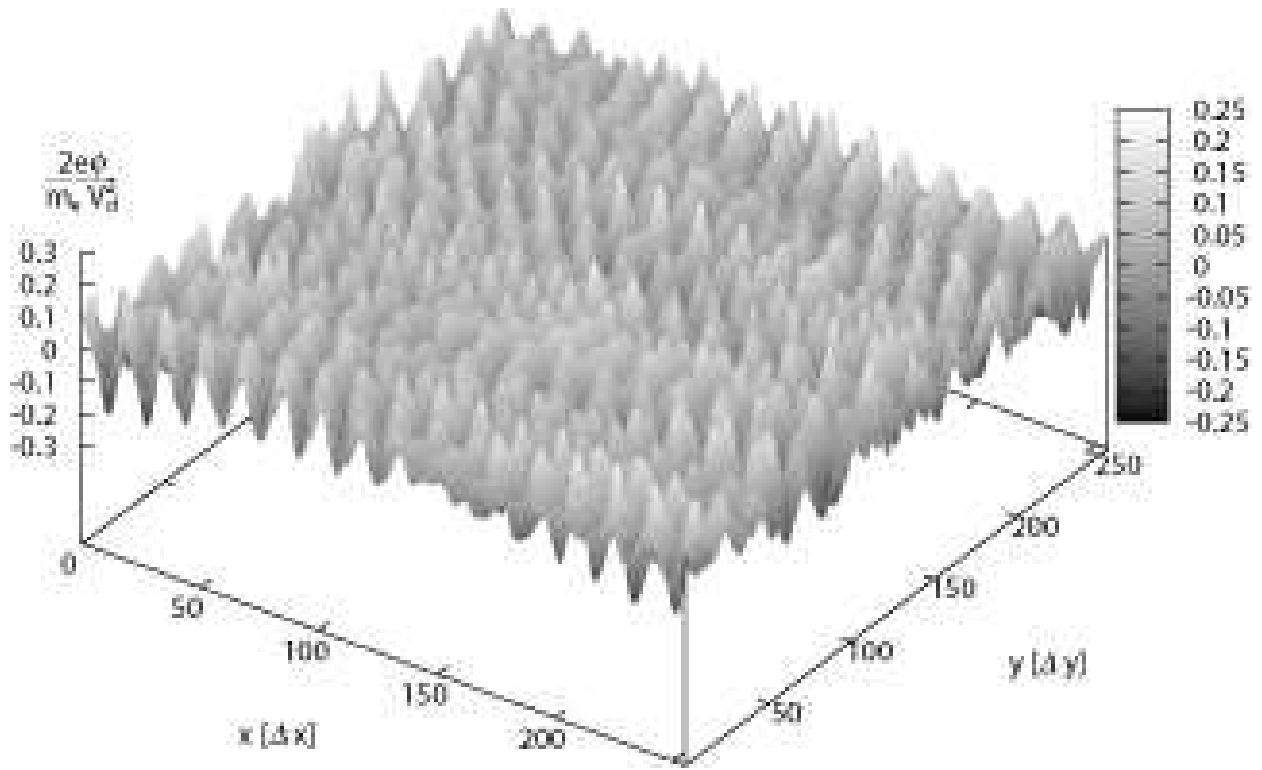} \caption{Landscape view of the electrostatic potential at t=150$\omega_{\rm pe}^{-1}$ in the 2D simulation. \label{fig2}}
\end{figure}


\begin{figure}
\plotone{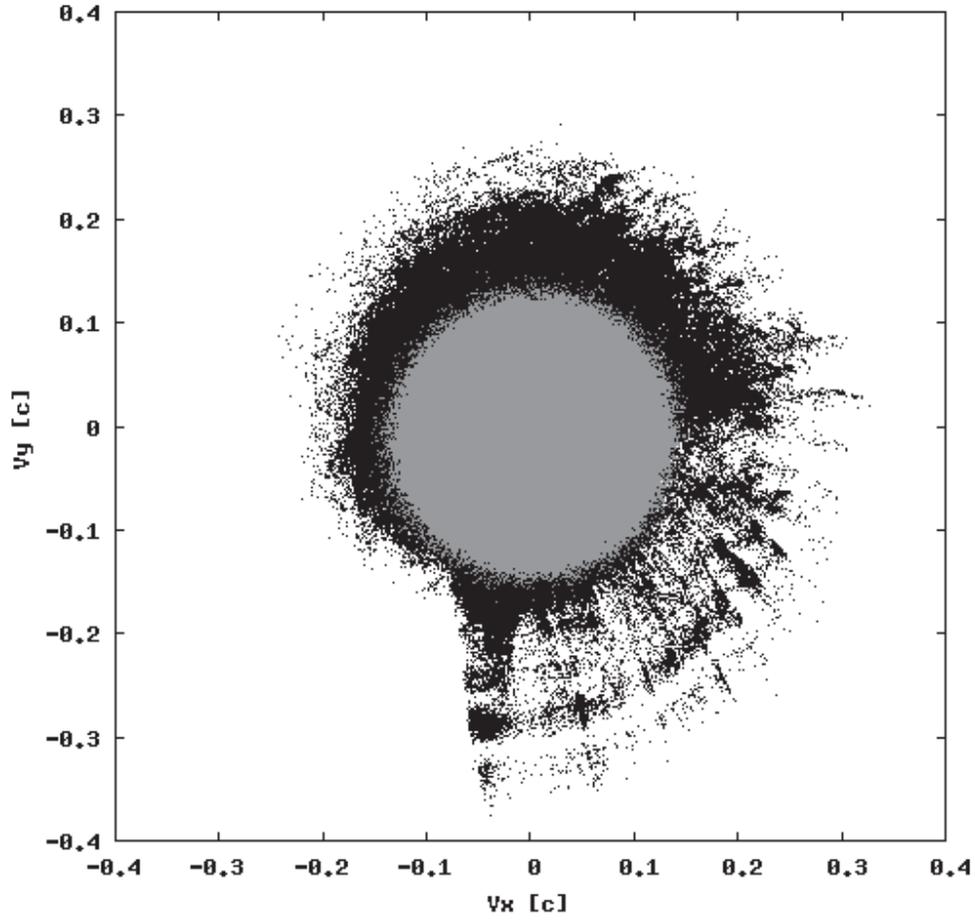} \caption{2D plot of the electron velocities at t=690 $\omega_{\rm pe}^{-1}$. Black and gray dots represent those of the 1D and 2D simulations, respectively. \label{fig3}}
\end{figure}


\begin{figure}
\plotone{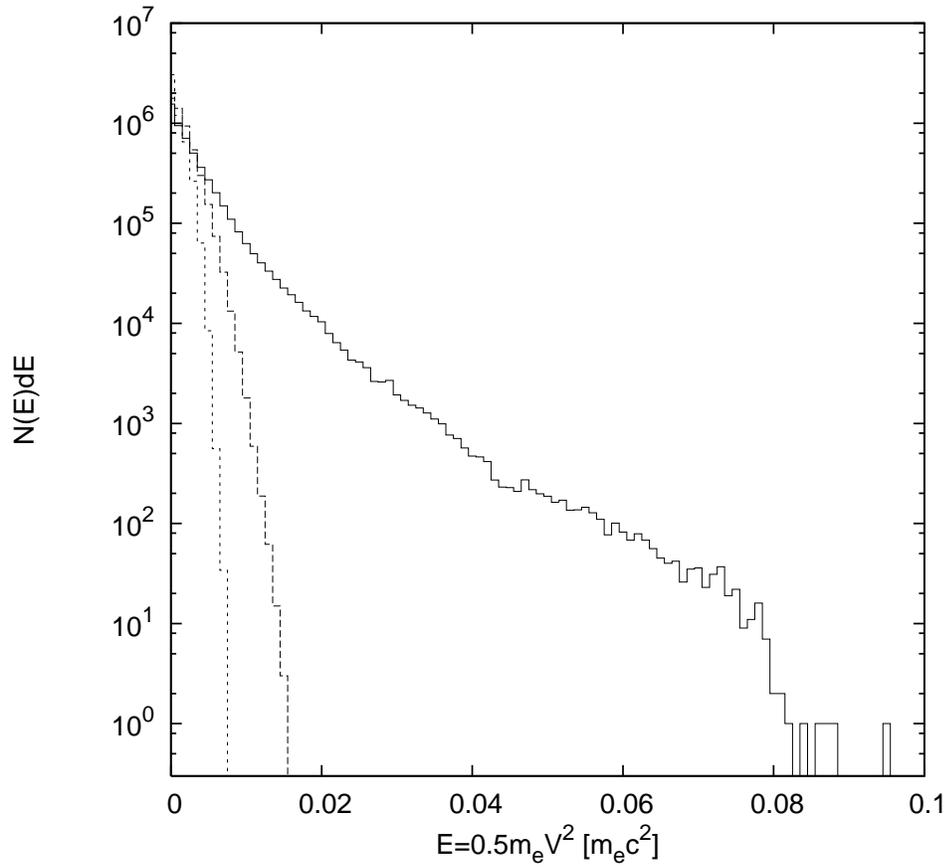} \caption{Energy spectrum of electrons at the final state of our simulation t=1000 $\omega_{\rm pe}^{-1}$. Solid, dashed and dotted histograms represent the 1D result, the 2D one
with magnetic field and the 2D one without magnetic field, respectively. \label{fig4}}
\end{figure}

\end{document}